# Charactering the magnetic properties of the copper chalcopyrite semiconductor CuGaSe2 via Monte Carlo simulations

**S. IDRISSI[1, *], N. EL MEKKAOUI[1], S. ZITI[2], H. LABRIM[3], R. KHALLADI[1], S. MTOUGUI[1], I. ELHOUSNI[1] and L. BAHMAD[1]**

[1]Laboratoire de la Matière Condensée et des Sciences Interdisciplinaires (LaMCScI), Mohammed V University in Rabat, Faculty of Sciences, B.P. 1014 Rabat, Morocco.

[2] Intelligent Processing and Security of Systems, Mohammed V University in Rabat, Faculty of Sciences, B.P. 1014 Rabat, Morocco.

[3] USM/DERS/Centre National de l'Energie, des Sciences et des Techniques Nucléaires (CNESTEN), Rabat, Morocco.

**Abstract:**

This manuscript presents a model and simulation of the copper chalcopyrite semiconductor CuGaSe2 in order to predict its magnetic properties. In the semiconductor material CuGaSe2 (CGS), the atom Cu is the only magnetic element with a magnetic moment value S = 1/2. We propose a Hamiltonian to calculate the energies corresponding to different configurations of the system. In a first step, we presented the magnetization behavior by using the Monte Carlo simulations (MCS). The magnetizations as a function of the temperature, the exchange coupling have been studied and discussed. We have also deduced and studied the magnetic susceptibilities. To complete this study, we have established and discussed the magnetic hysteresis loops of the studied compound. Also we have calculated the corresponding critical exponents and compared them with the standard 3D Ising model.

KEYWORDS:

Semiconductor; Copper chalcopyrite CuGaSe2; Monte Carlo; Hysteresis loops; critical Exponents.

*) Corresponding authors: lahou2002@gmail.com; samiraidrissi2013@gmail.com

## I. Introduction

Recently, the family of the Chalcopyrite semiconductors such as: CuInSe2 (CIS), CuGaSe2 (CGS) are useful compounds for efficient thin film solar cells [1]. These materials show interesting optical properties for solar-cell application. However, the band gap energy of the compound $CuGaSe_2$ (1.68 eV) is high when compared to the predicted optimal band gap 1.4 eV [2]. This value is also situated above the band gap energy through the Sommerfield effect [3]. On the other hand, the researches existing in the literature show some difficulties in the electrical transport characteristics in CuGaSe2. It has been found that the grain boundaries of CuInSe2 and CuInGaSe2 do not produce fatal leakage paths in the solar-cell device applications. Because of the energetic structure of cation terminated grain boundaries of this compound, this beneficial effect is not seen in the material CuGaSe2 [4].

It has been found that the undoped CuGaSe2 electrical transport shows high concentration hole, because of the Cu vacancy. The formation energy of Cu vacancy is found to decrease when the Fermi energy increases. Also, the Fermi energy increases at a certain level because of the compensating acceptor concentration, caused by the Cu-vacancy increases with Fermi energy. Concerning the compound CuInSe2, a narrow bandgap materials such as CuInSe2, shows the n-type conductivity and could be achieved due to n-type conversion [5]. However, Ge-doped n-type CGS has been studied and it reports ion implantation on bulk CuGaSe2 [6]. Efficient donor centers in ternary CuGaSe2 are produced by Group-IV elements such as C, Si, Ge and Sn. The element Si as the donor impurity has been first studied in experimental studies with the group-IV elements. The formation of donors is confirmed through photoluminescence (PL) measurement. Besides, the donor-acceptor pair emission intensity gains considerably by increasing Si doping in CuGaSe2. Up to room temperature, ternary semiconductors alloys, with magnetic impurities, have attracted substantial attention, because of their presumed ferromagnetism. The appealing optical properties of pure chalcopyrites and their characteristics are promising photovoltaic devices [7]. Palacios et al. [8, 9] have suggested doped chalcopyrites transition metal applications. On the other hand, Han et al. [10] have outlined that Fe and Ni doped CuGaS2 could be useful materials for IBSC applications. Transition metals doped were investigated by Wang et al. [11], and they found that Ti doped CuAlS is the most promising candidate. When doping $CuGaSe_2$ with Ti [12], Sn [13] and Fe [14], photovoltaic devices have been synthesized and based on Fe.

Absorbers with a band-gap above 1.14 eV are good candidates to host an intermediate band [15]. It is shown that $CuGaSe_2$ has a relatively wide band-gap of 1.68 eV [16]. Moreover, various types of the compounds have been examined by Monte Carlo simulations [17-30]. The purpose of this article is the study the magnetic properties of the compound $CuGaSe_2$ by Monte Carlo simulations. This work presents a model to simulate the magnetic properties of this compound. We start this study by proposing a Hamiltonian, see Eq. (1) to calculate magnetic properties of this compound. In fact, we study the magnetizations and magnetic susceptibilities by using the Monte Carlo simulations (MCS). The effect of varying different physical parameters has been presented and discussed, such as the exchange coupling interactions and the external magnetic field. We have also studied and discussed the magnetic hysteresis loops of the compound $CuGaSe_2$. We have calculated the corresponding critical exponents of this compound.

## II. Structure of the model:

The family ternary chalcopyrite type crystal structure with the chemical formula I-III-VI2 such as CuGaSe2 space group $I\bar{4}2d$ is similar to the mono-elemental diamond. The ternary compound CuGaSe2 has a chalcopyrite structure. It is a diamond-like structure similar to the sphalerite structure, but with an ordered substitution of group I (Cu) elements and those of group III (Ga or In) to group II (Zn) elements. CuGaSe2 is a ternary compound of type I-III-VI2. It has the same structure as the CuInS. It has two allotropic forms: Sphalerite and Chalcopyrite. The elementary cell is shown in Fig.1a. Thus each selenium atom is tetraedrically bound to two atoms of copper and two atoms of gallium, each of them surrounded by four atoms of selenium. The crystalline parameters of CuGaSe2, according to several authors, are shown in Table 1, we note that the degree of tetragonality expressed by the ratio c/a is less than 2.

***Table 1: Parameters of the compound CuGaSe2.***

| a (Å) | c (Å) | c/a | Ref. |
|---|---|---|---|
| 5.614 | 11.03 | 1.965 | [22] |
| 5.5963 | 11.0036 | 1.966 | [23] |

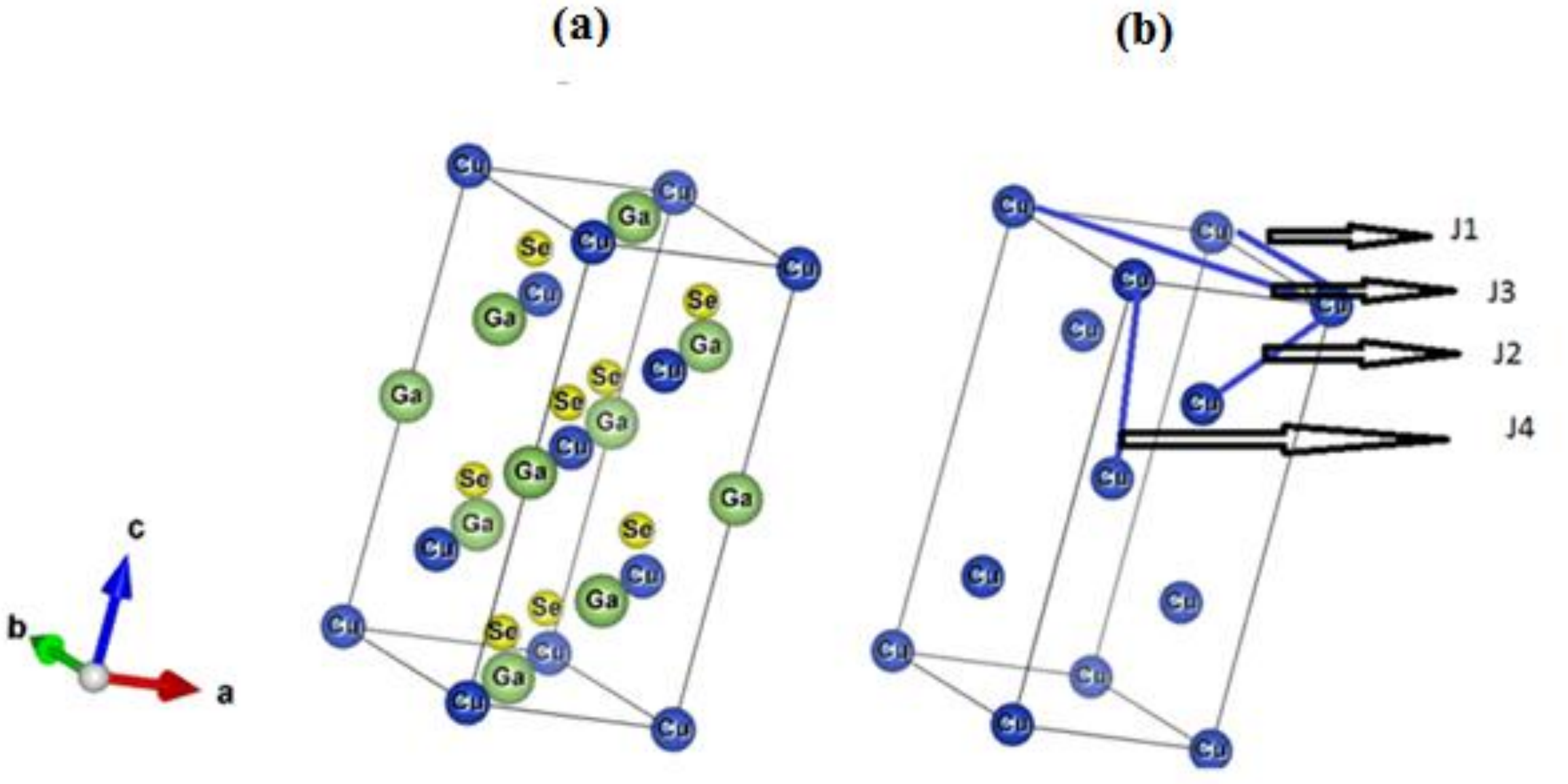

***Fig.1: (a) Unit cell of the CuGaSe2 structure. (b) Structure of the compound CuGaSe2 describing the different exchange couplings.***

The geometry of the studied system is belonging to the space group I4¯2d (No. 122) and is displaying in Fig. 1, using VESTA software [31]. The Cu sublattice has spins S and takes the values ± 1/2.

The Hamiltonian controlling this system takes is as following:

$$\mathcal{H} = -J1\sum_{<i,j>} S_i\, S_j - J2\sum_{<i,k>} S_i\, S_k - J3\sum_{<i,l>} S_i\, S_l - J4\sum_{<k,l>} S_k\, S_l - H\sum_i S_i \quad (1)$$

Where: $S_i=\pm 1/2$,

$J_1= J_{Cu\text{-}Cu}$ describes the exchange interaction between the first neighbor Cu atoms.

$J_2= J_{Cu\text{-}Cu}$ is the exchange interaction between the second neighbor Cu atoms.

$J_3= J_{Cu\text{-}Cu}$ stands for the exchange interaction between the third neighbor n Cu atoms.

$J_4= J_{Cu\text{-}Cu}$ stands for the exchange interaction between the fourth neighbor Cu atoms.

H is the external magnetic field.

## III. Model and formulations:

The Monte Carlo simulation (MCS) under the Metropolis algorithm, have been used to simulate the magnetic properties of the compound CuGaSe2 [32]. This method is found to be most efficient when investigating the magnetic behavior of several complex structures, always difficult to be solved with other simulation methods. The Metropolis algorithm allows the acceptation or rejection of generated configurations. The energy and the magnetizations are calculated. Then, we deduce the average values of the physical quantities, at equilibrium. Such physical quantities are: the magnetic susceptibility $\chi$ and the specific heat $C_v$. The equations used to calculate these parameters are:

$$E = \frac{1}{N} < \mathrm{H} > \qquad (2)$$

$$m = \frac{1}{N} < \sum_i Si > \qquad (3)$$

$$\chi = \beta(< m^2 > - < m >^2) \qquad (4)$$

$$C_v = \beta^2(< E^2 > - < E >^2) \qquad (5)$$

Where $\beta = \frac{1}{k_B T}$ and $k_B$represents the Boltzmann constant. For simplicity, we fixe $k_\mathrm{B} = 1$, in all the following.

The critical behaviors of the observable quantities are determined in [33–36]. We introduce the reduced temperature: t= (T—Tc)/Tc, where Tc is the critical temperature, so that the critical parameters: α, β and γ are defined as follows:

$$C_V \sim t^{-\alpha} \qquad (6)$$

$$m \sim (-t)^{\beta} \qquad (7)$$

$$\chi \sim t^{-\gamma} \qquad (8)$$

Where the parameter α is deduced from the behavior of the specific heat Cv, the parameter β is related to the order parameter (magnetization) and γ corresponds to the behavior of the susceptibility $\chi$ of the studied copper chalcopyrite semiconductor CuGaSe2 compound

## IV. Results and discussion

In this section, we will examine the magnetic properties in case of the T > 0 K. We study the magnetic properties of sulfide CuGaSe2 using Monte-Carlo simulation based on the Metropolis algorithm. The principle of this simulation is to change the value and/or the direction of the spin of all sites in the lattice for every Monte Carlo step. We study the magnetic properties of the copper chalcopyrite semiconductor CuGaSe2 (CGS), using Monte Carlo simulations. This computation method is based on the Hamiltonian given in Eq. (1). In this study, the free boundary conditions on the lattice are used. In this study we provide numerical values for the specific super-cell size N=5x5x5. For every spin Cu (S=1/2) configuration, $10^5$ Monte Carlo steps are performed. All the sites of the system are visited and a single-spin flip attempt is made. Each flip is accepted or rejected according to the probability based on the Boltzmann statistics under the Metropolis algorithm.

To study the thermal behavior of the physical parameters of the compound CuGaSe2, we plot in Fig.2a the corresponding magnetization and the susceptibility in Fig.2b. The peak of susceptibility is located at critical temperature Tc= 4 K.

In Fig.3 we plot the thermal behavior of the energy of the compound CuGaSe2 for H=0, $J_1$=$J_2$=$J_3$=1 and $J_4$ =1.414. From this figure, the total energy of the system increases when increasing the temperature of the system. In order to study the behavior of the specific heat of this compound we plot in Fig.4 the obtained results. It is found that this physical parameter exhibits a maximum in the range of T=4 K. This is a good agreement with the behavior of the susceptibility shown in Fig.2b.

We provide in Fig.5 the hysteresis loops of the studied system for selected values of the exchange coupling $J_1$ and $J_2$: (J1=J2=J3=1 and J4=1.414), for different values of the Temperature (T=0.5, T=1.5 and T=3). From this figure, it is found that the coercive field decreases when increasing the temperature. In fact the surface of the hysteresis loops becomes null for temperatures T > 5 K.

To complete this study, we plot in Figs. 6a, 6b and 6c, the behavior of the total magnetizations as a function of the different exchange coupling interactions: J1, J2 and J3, respectively. In fact, Fig.6a provides the magnetizations as a function of exchange coupling J1 for fixed values of J2=J3=1 and J4 =1.414. Also, in this figure we fixed H=0 and selected values of temperature T=0.5 K, T=1 K and T=2 K. For negative and large values of J1, the total magnetizations are not affected by the variations of this parameter. For positive and large values of J1, the magnetization saturation is reached. From this figure, when varying the temperature and increasing the exchange coupling interaction J1, the magnetization saturation is reached either positively or negatively.

To inspect the effect of the variation the parameter J2 on the total magnetizations, we plot in Fig. 6b the corresponding results for: H=0, J1=J3=1, J4 =1.414 and selected values of temperature T=0.5 K, T=1 K and T=2 K. From this figure, it is found that when varying the exchange interaction between Cu atoms J2, the total magnetizations are by the increasing temperature effect, from T=0.5 K to T=2 K, see Fig. 6b. The magnetization saturation is reached for positive and large values of J2. To complete this study, we illustrate in Fig. 6c, the total magnetization as a function of exchange coupling J3, for H=0, J1=J2=1, J4 =1.414 and selected values of temperature T=0.5 K, T=1 K and T=2 K. otherwise the negative values of exchange coupling J3, the saturation is reached when increasing the parameter J3.

In order to investigate the critical exponents of the studied compound CuGaSe2, formulated by equations (6, 7 and 8), we provide in Figs.7, 8 and 9, the corresponding results. In fact, Fig. 7 summarizes the behavior of log (m) versus log (Tc-T)/Tc for the compound CuGaSe2. The slope of this figure represents the critical exponent value $\beta$=0.150. Also, we plot in Fig. 8, the behavior of log ($\chi$) versus log (T-Tc)/Tc of the studied compound CuGaSe2. From this figure, the calculated slope is the critical exponent $\gamma$=1.20. Regarding the specific heat, we illustrate In Fig. 9 the behavior of log (Cv) versus log (T-Tc)/Tc .The obtained slope is the critical exponent $\alpha$=4.00. This is in a good agreement with values in the literature of the 3D ising model, see Ref. [37] and references therein. Preliminary calculations of the critical exponents have been carried out for other system sizes: 3x3x3, 4x4x4, 5x5x5, 6x6x6, … etc; showing that these parameters did not change in a significant way from the system size 5x5x5. For this reason, we have fixed the system size 5x5x5 for our calculations in this work. Comparing our results with the standard 3D Ising model, we found that the all critical exponents coefficients are closer.

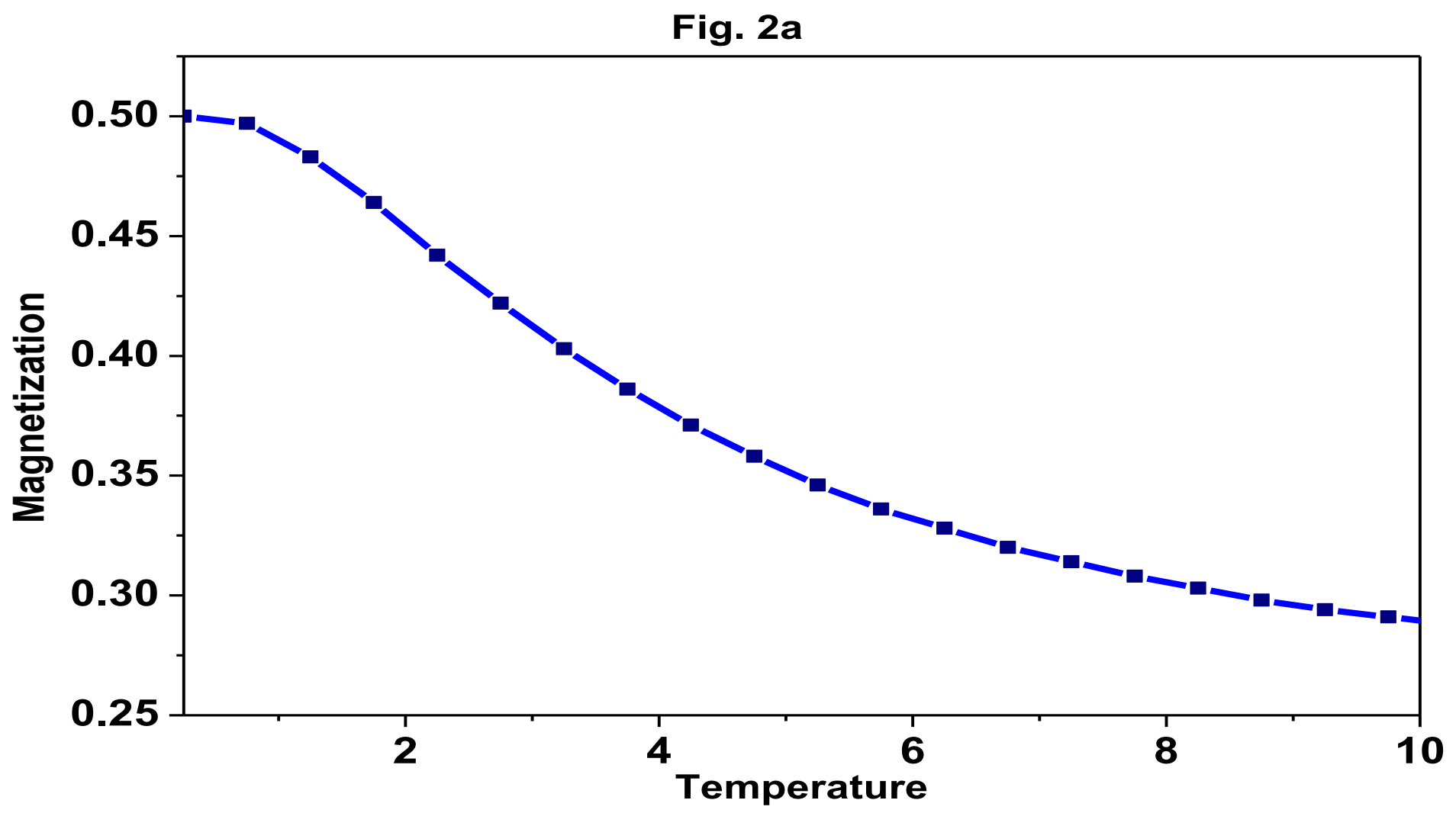

*Fig.2a. Thermal behavior of the total magnetization for H=0, J1=J2=J3=1 and J4 =1.414.*

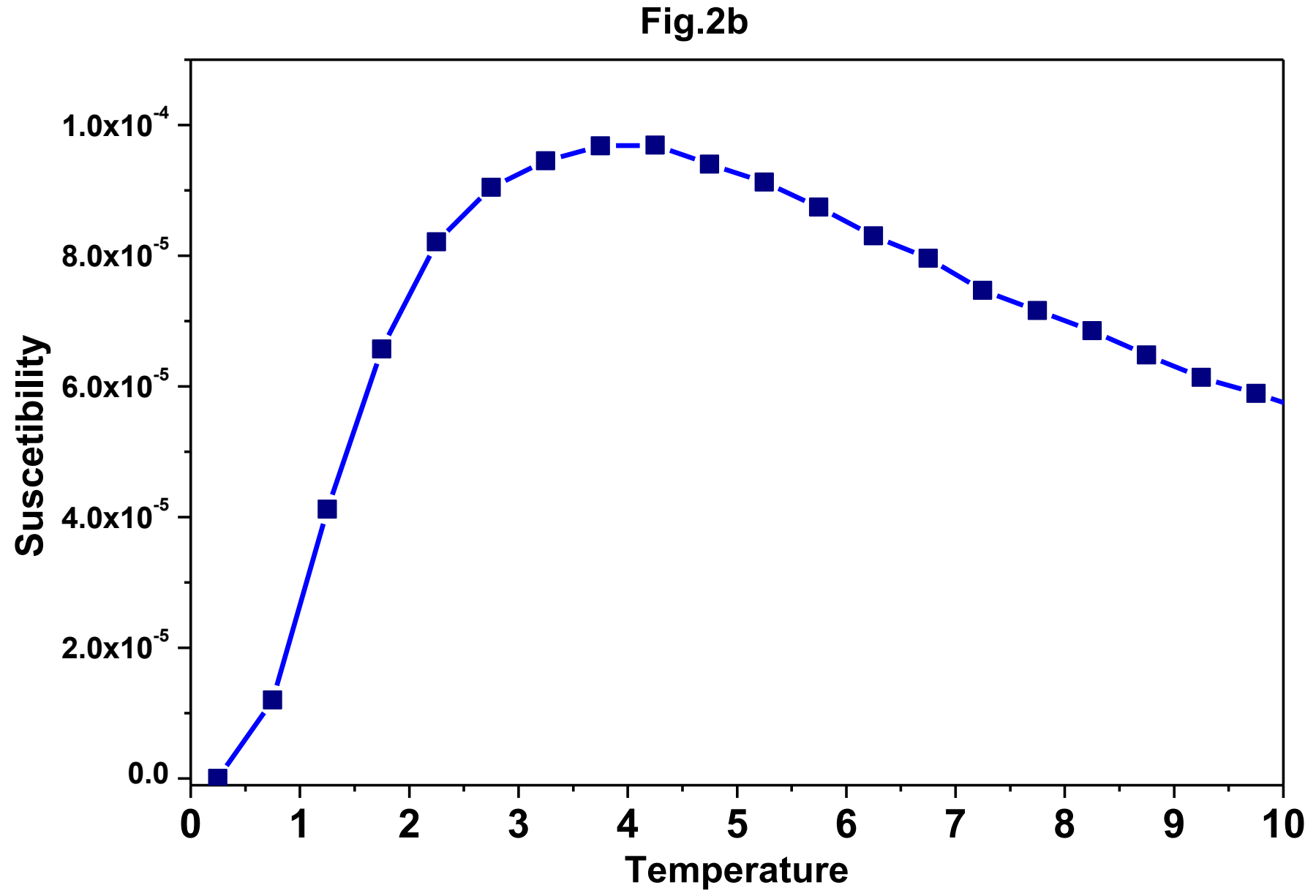

*Fig.2b. Thermal behavior of the total susceptibility for H=0, J1=J2=J3=1 and J4 =1.414.*

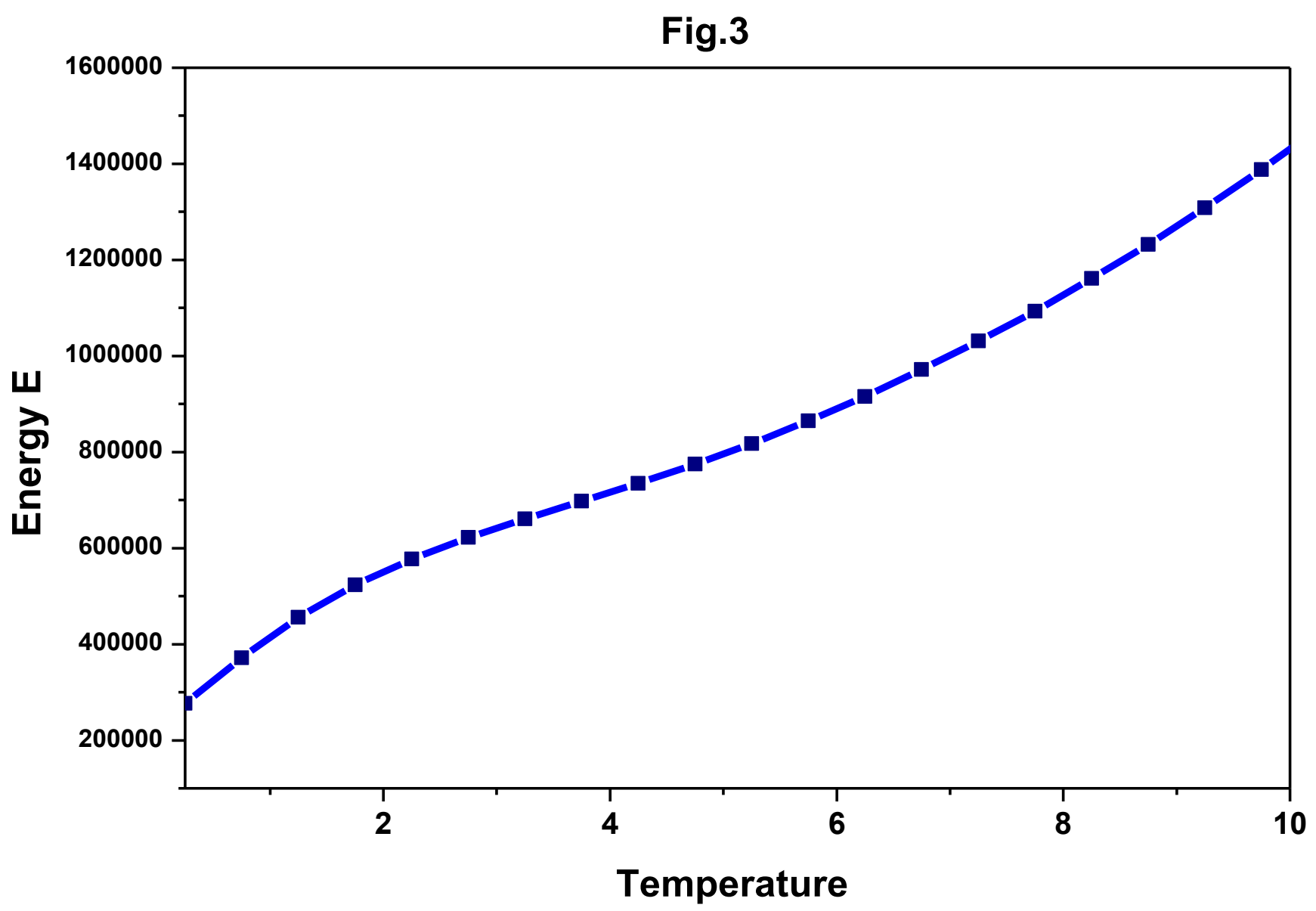

*Fig. 3. Thermal behavior of energy of the system for H=0, J1=J2=J3=1 and J4 =1.414.*

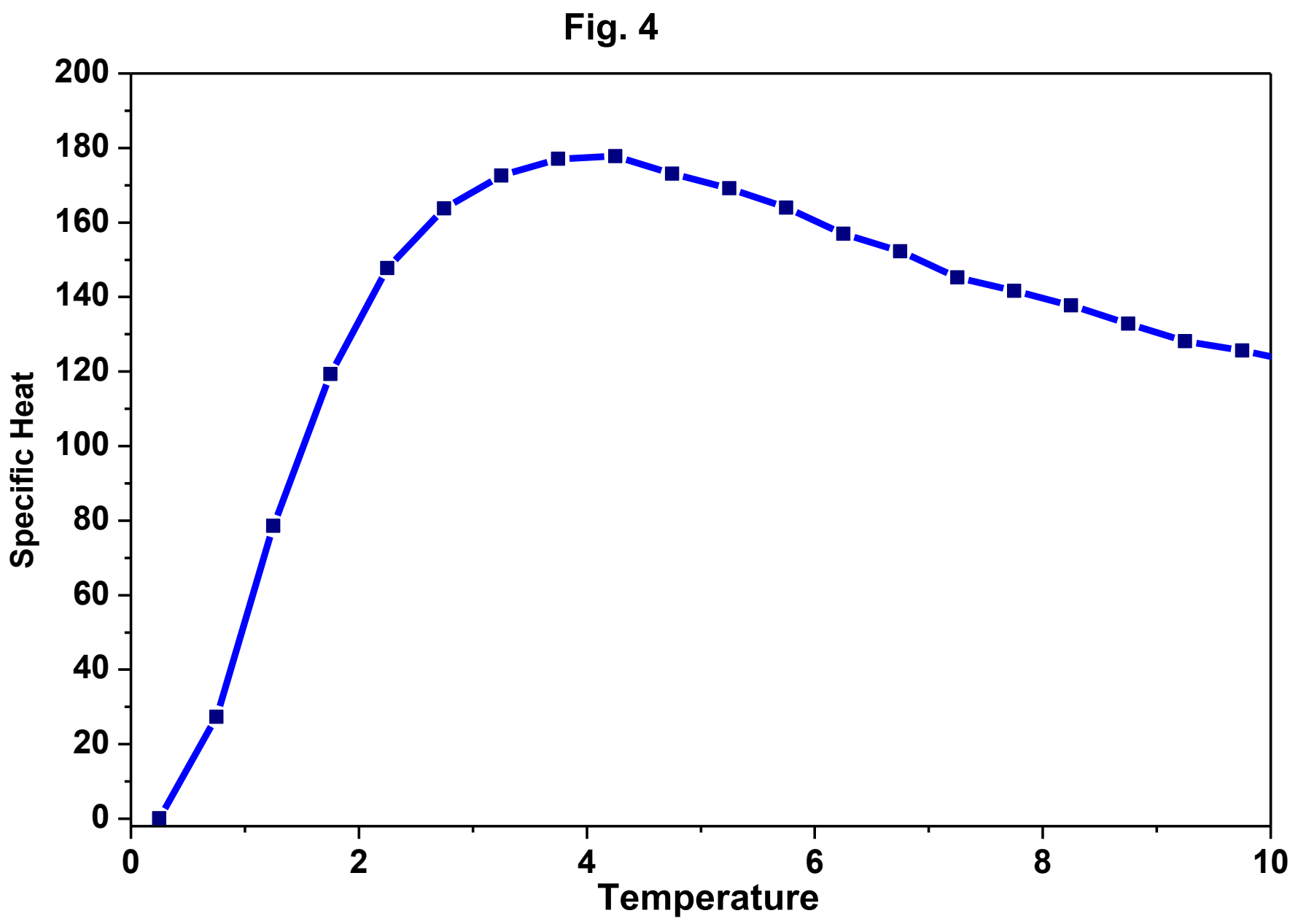

*Fig. 4. Thermal behavior of the specific heat for $H=0$, $J_1=J_2=J_3=1$ and $J_4=1.414$.*

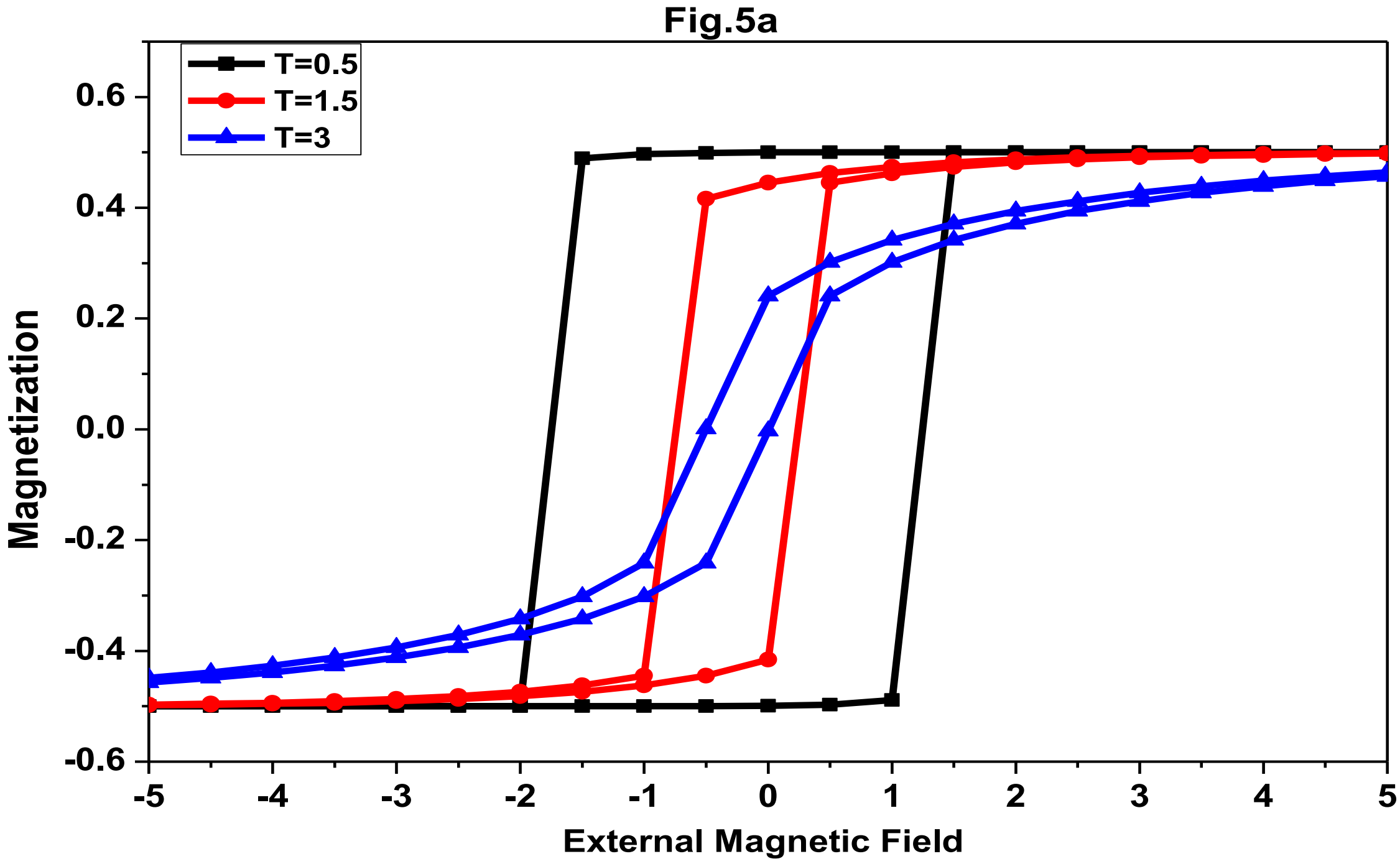

*Fig.5. Hysteresis loops of the system CuGaSe2 for selected values of the exchange coupling J1 and J2 :(J1=J2=J3=1 and J4=1.414) and different values of the Temperature :T=0.5, T=1.5 and T=3.*

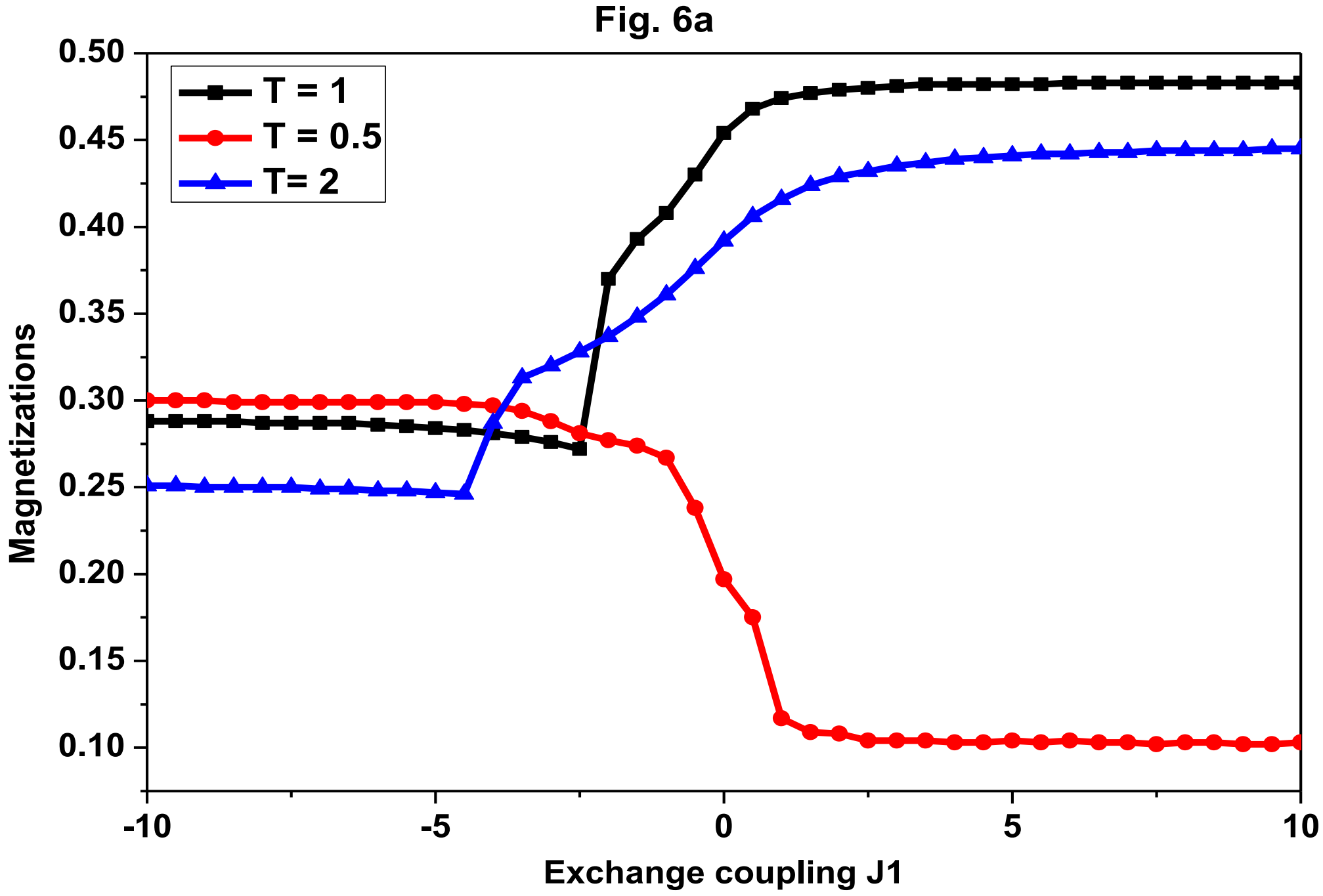

*Fig. 6a. Magnetization as a function of exchange coupling J1 for J2=J3=1 and J4 =1.414, H=0 and selected values of temperature T=0.5 K, T=1 K and T=2 K.*

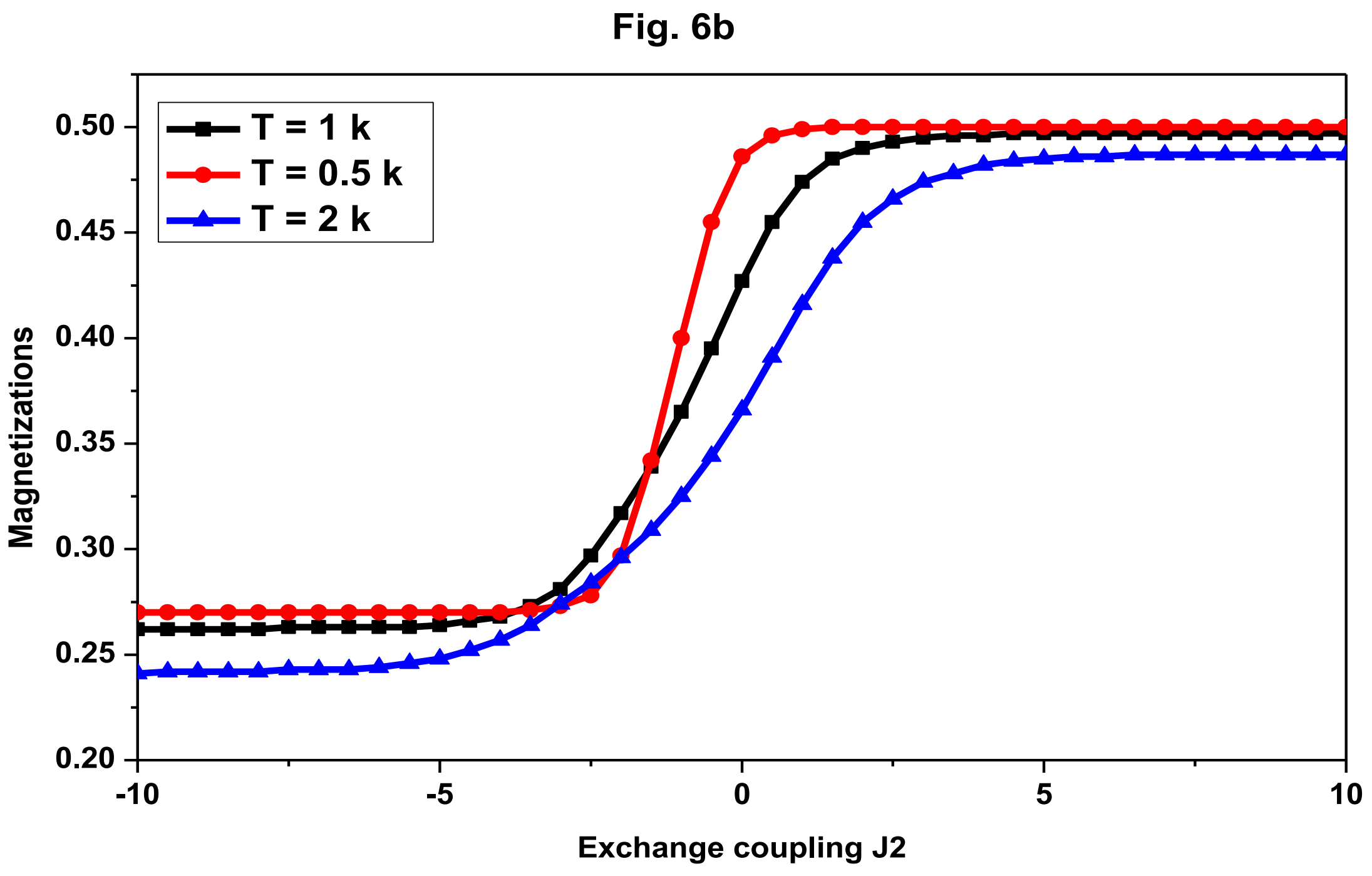

*Fig. 6b: Magnetization as a function of exchange coupling J2 for J1=J3=1 and J4 =1.414, H=0 and selected values of temperature T=0.5 K, T=1 K and T=2 K.*

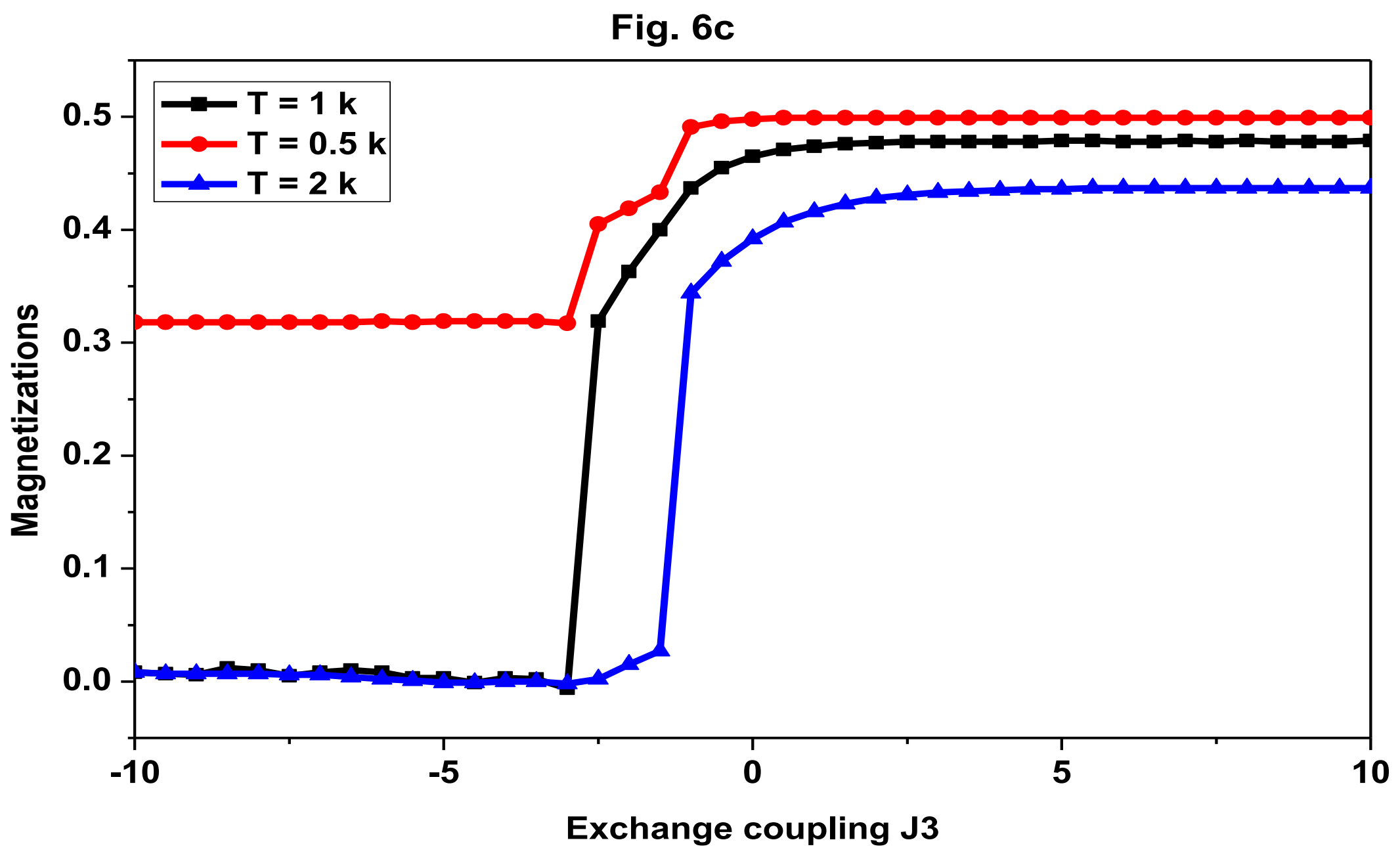

*Fig. 6c. Magnetization as a function of exchange couplings: J3 for J1=J2=1 and J4 =1.414, H=0 and selected values of temperature T=0.5 K, T=1 K and T=2 K.*

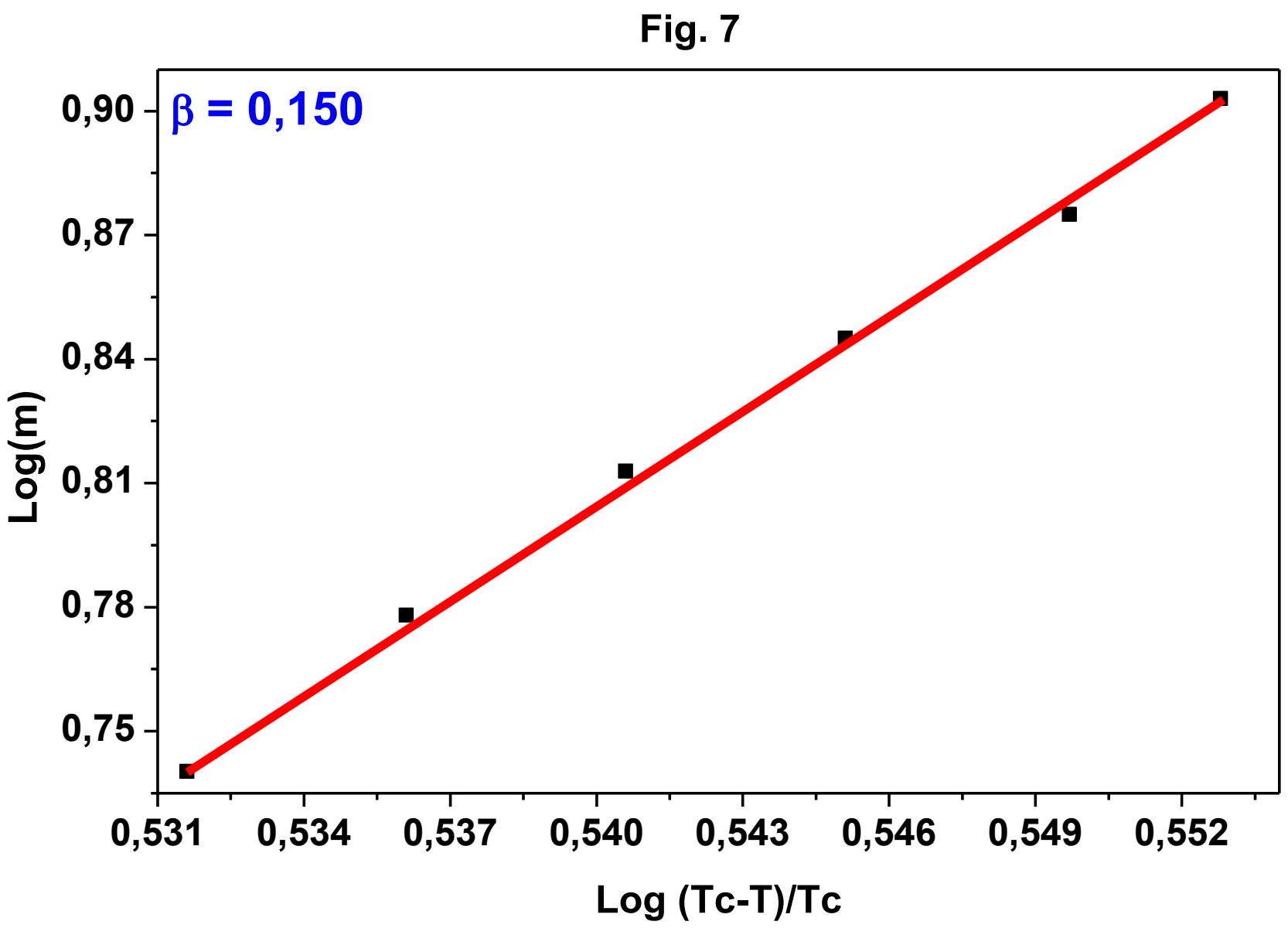

*Fig. 7: The behavior of log (χ) versus Log (Tc-T)/Tc of the compound CuGaSe2. The slope represents the critical exponent β.*

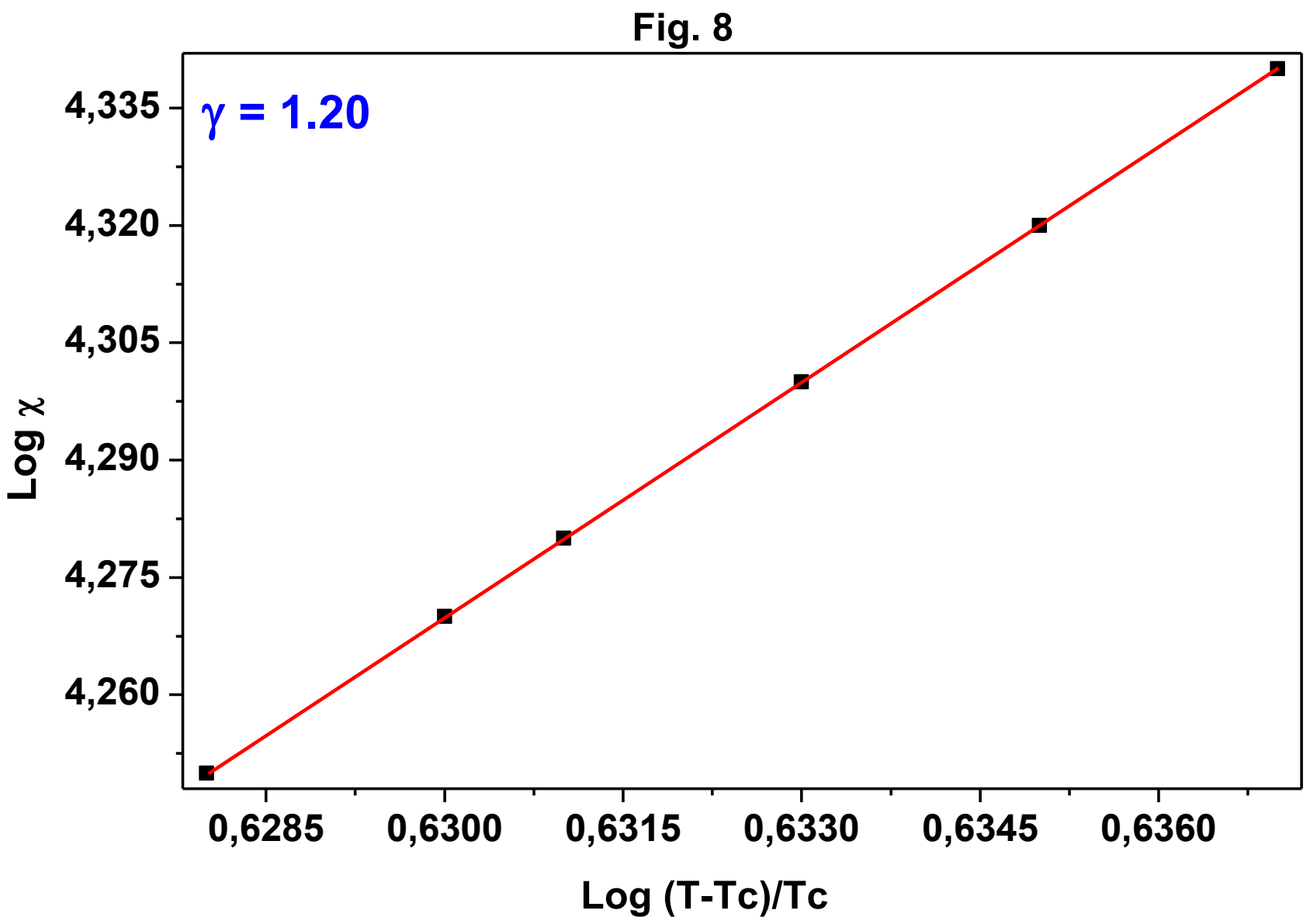

*Fig. 8: The behavior of log (χ) versus Log (T-Tc)/Tc of the compound CuGaSe2. The slope represents the critical exponent γ.*

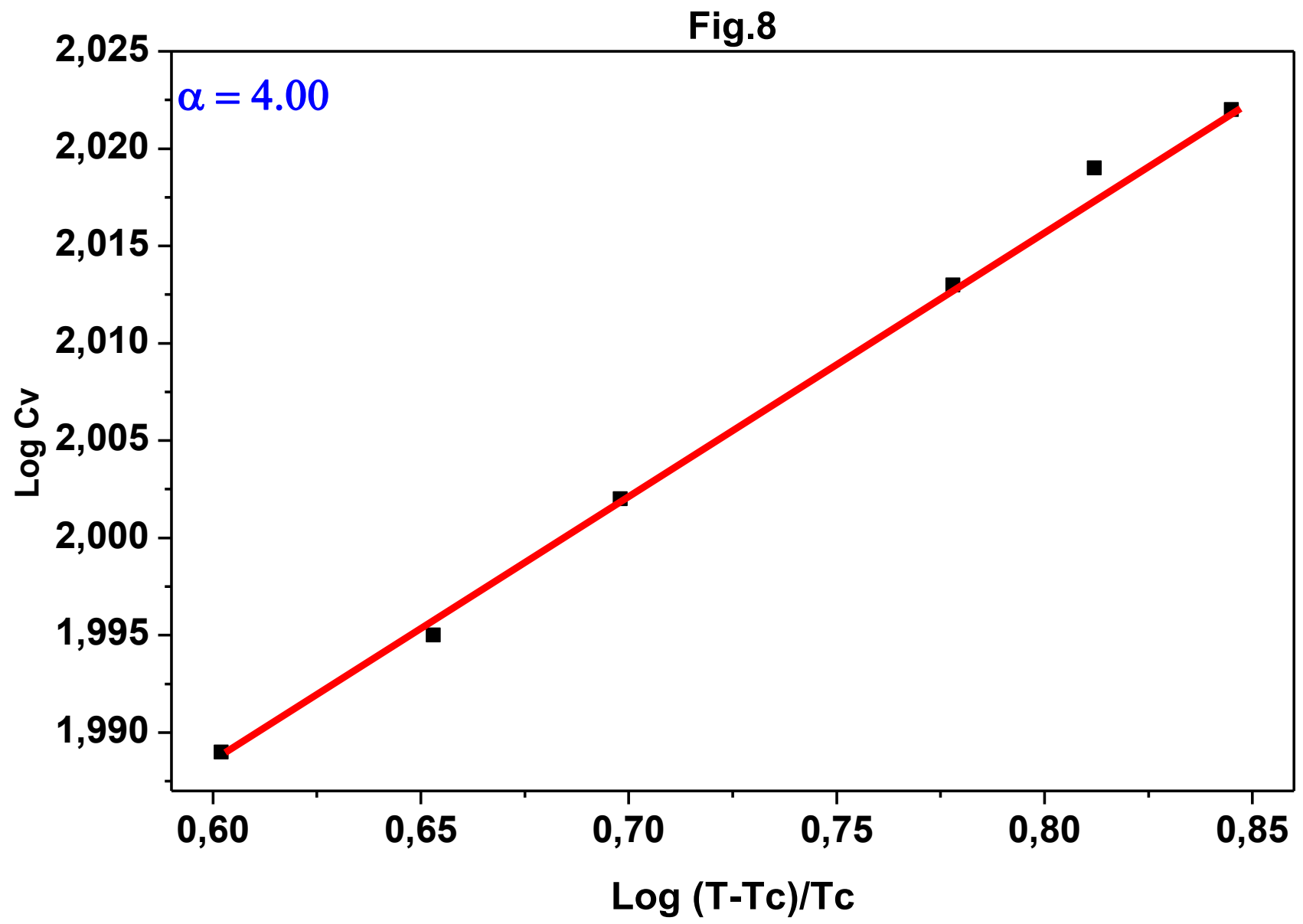

***Fig. 9: The behavior of log (Cv) versus Log (T-Tc)/Tc of the compound CuGaSe2. The slope represents the critical exponent α.***

## V. CONCLUSION

In this work, we have studied and simulated the magnetic properties of copper chalcopyrite material CuGaSe2. For this purpose, we used the Monte Carlo simulations (MCS) based on Metropolis algorithm. We proposed a Hamiltonian to describe the studied system and calculate the magnetization profiles of this compound. In fact, we have presented the obtained results of the CuGaSe2 compound. The magnetizations and the susceptibilities as a function of the temperature, and the exchange coupling interactions have been deduced and discussed. Moreover, the effect of varying different exchange coupling interactions is presented for specific values of the physical parameters. To complete this study we have presented the magnetic hysteresis loops for selected values of the temperature, and the exchange coupling interactions. Also we have calculated the corresponding critical exponents and compared them with the standard 3D Ising model.